\documentclass[twocolumn,prl]{revtex4}   
\usepackage{graphicx}
\usepackage{CJK}

\begin{document}

\title{Triply-resonant Optical Parametric Oscillator by Four-wave Mixing with Rubidium Vapor inside an Optical Cavity}

\author{Xudong Yu$^{1}$, Min Xiao$^{2}$, and Jing Zhang$^{1\dagger}$}

\affiliation{$^{1}$The State Key Laboratory of Quantum Optics
Quantum Optics Devices, Institute of Opto-Electronics, Shanxi
University, Taiyuan 030006, P.R. China}

\affiliation{$^{2}$ Department of Physics, University of Arkansas,
Fayetteville, Arkansas 72701, USA}

\begin{abstract}
We present an experimental demonstration of simultaneous
above-threshold oscillations of the Stokes and anti-Stokes fields
together with the single pumping beam with rubidium atoms inside an
optical standing-wave cavity. The triple resonant conditions can be
achieved easily by making use of the large dispersions due to
two-photon transitions in the three-level atomic system. This work
provides a way to achieve high efficient nonlinear frequency
conversion and the generated bright Stokes and anti-Stokes cavity
output beams are potential resource for applications in quantum
information science.

\end{abstract}

\maketitle

Various atomic systems have been used as intracavity gain media to
realize cavity oscillations
\cite{one,two,three,four,five,six,seven}. Especially, multi-level
atomic systems have more interesting characteristics and can be more
efficient in managing the absorption and dispersion properties of
the intracavity gain medium for building up resonance simultaneously
of different cavity modes. It is easy to have an off-resonant pump
beam to make the Stokes or anti-Stakes field be on resonance with
one of the cavity modes in a three-level $\Lambda$-type atomic
system, and get it to oscillate with a large pump power
\cite{seven}. When the pump beam is tuned near one of the atomic
transitions in a three-level $\Lambda$ system inside an optical
cavity, cavity field oscillation (or lasing without population
inversion) can occur in the frequency corresponding to the other
atomic transition \cite{Wu}. This system can be considered as an
optical parametric oscillator (OPO) since the atomic variables can
be adiabatically eliminated in treating the atom-field interactions
\cite{Guzman,Xiong}.

In this Letter, we present our experimental demonstration of a
triply-resonant atom-cavity system, as shown in Fig.1. With one pump
laser beam (frequency $\omega_{p}$) tuned to the cavity resonance as
the cavity input, two optical fields, both the Stokes field
(frequency $\omega_{b}$) and anti-Stokes field (frequency
$\omega_{a}$), are generated simultaneously. The frequency of the
pump beam is detuned from the atomic resonances
$|1\rangle\leftrightarrow|0\rangle$ and
$|2\rangle\leftrightarrow|0\rangle$ by the amount of
$\Delta_{b}=\omega_{p}-\omega_{01}$ and
$\Delta_{a}=\omega_{p}-\omega_{02}$ (thus
$\Delta_{a}-\Delta_{b}=\omega_{12}$), which generates two sidebands
at frequencies of $\pm\omega_{12}$ from the pump beam frequency,
respectively. Typically, the generated Stokes and anti-Stokes fields
are difficult to be made on resonance with the cavity modes at the
same time, since the cavity mode has already been tuned to be on
resonance with the pump field. However, since we work with the
naturally mixed $^{87}Rb$ and $^{85}Rb$ vapor cell as the
intracavity medium, there are several broad absorption bands in the
transmission spectrum, as shown in Fig. 2(a). The weak field cavity
transmission spectrum is given in Fig. 2(b), which shows the cavity
transmission peaks with modified peak separations caused by the
enhanced dispersions associated with the tails of the absorption
lineshape at high atomic density \cite{Yu}. By manipulating the
large intracavity dispersions via pump laser frequency detuning, the
Stokes and anti-Stokes fields can be made to be simultaneously on
resonance with the modified cavity modes together with the pump
field, which is very similar to the case of a triple-resonant OPO
with nonlinear crystals
\cite{Fabre,Villar,Villar1,Villar2,Coelho,Dauria,Su,Laurat}.

%
\begin{figure}
\centerline{
\includegraphics[width=3in]{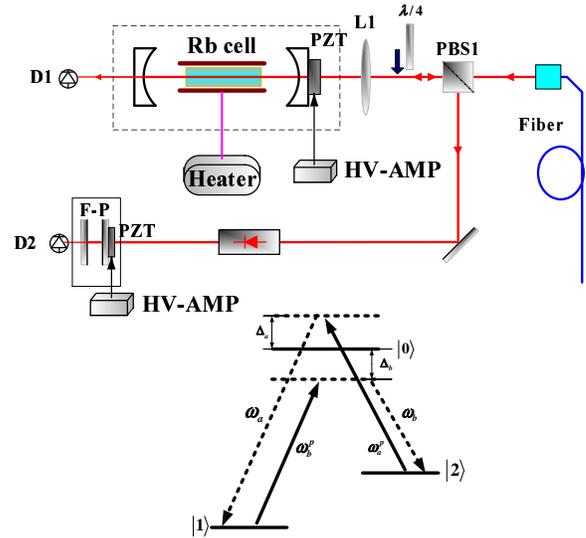}
} \vspace{0.1in}
\caption{ (Color online). Schematic of the experimental setup of the
coupled three-level atoms-cavity system. $\lambda /4$: quarter-wave
plate; D1, D2: detectors; HV-AMP: high voltage amplifier; PZT:
piezoelectric transducer; PBS: polarized beam splitter; $L$: optical
lens; F-P: Febry-Perot cavity. Bottom: relevant energy levels and
field configurations. $\omega_{a}^{p}=\omega_{b}^{p}=\omega_{p}$ is
the frequency of the single pump field. \label{Fig1} }
\end{figure}

The experiment is done by placing a naturally mixed rubidium vapor
cell inside an optical standing-wave cavity of length 17.7 $cm$. The
cavity is composed of two curved mirrors with the same radius of
curvature of 100 $mm$. The reflectivity is $90\%$ at 780 $nm$ for
the input coupler $M1$ (on right), which is mounted on a PZT to
adjust the cavity length. The left mirror $M2$ has a reflectivity of
$99.5\%$ at 780 $nm$. The finesse of the cavity (including the
losses of two faces of the atomic cell) is about $F=20$. The length
of the vapor cell is 7.5 $cm$. Thus we may obtain that the cavity
bandwidth (the half-width at half maximum for the cavity) is about
21 $MHz$ and the OPO efficiency (output coupling over total losses)
about $20\%$. The temperature of the vapor cell can be controlled by
a heater. A beam from a grating-stabilized diode laser is injected
into a tapered amplifier (TA). The high power output from the TA
then passes a standard polarization maintaining single-mode fiber,
which is used as the cavity pump beam with an input power of 100
$mW$. The pump laser beam with the spatial mode filter by the
optical fiber is easy to be mode-matched to the $TEM_{00}$ mode of
the optical cavity. This atom-cavity system is studied by monitoring
the cavity reflection spectra using a scanned F-P cavity. We explore
two different input and output configurations of the cavity to
detect the different polarizations of the generated Stokes and
anti-Stokes fields. One is that the pump field first passes through
a polarized beamsplitter (PBS) and is injected into the cavity with
horizontal polarization. The vertically-polarized component of the
cavity reflection field is reflected by the same PBS, which mainly
contains the generated Stokes and anti-Stokes fields with only a
little pump field. The total reflected field then passes through an
optical isolator and is monitored by a scanned F-P cavity. The other
configuration is to use the pump field to inject into the cavity
with circular polarization by passing through a $\lambda/4$
waveplate. The output pump field from cavity reflection is reflected
by the PBS when passes through the $\lambda/4$ waveplate again.
Thus, the total reflected field from the PBS contains the reflected
pump field from the cavity and also the generated Stokes and
anti-Stokes fields with the same polarization as the pump field
inside the cavity.

%
\begin{figure}
\centerline{
\includegraphics[width=3in]{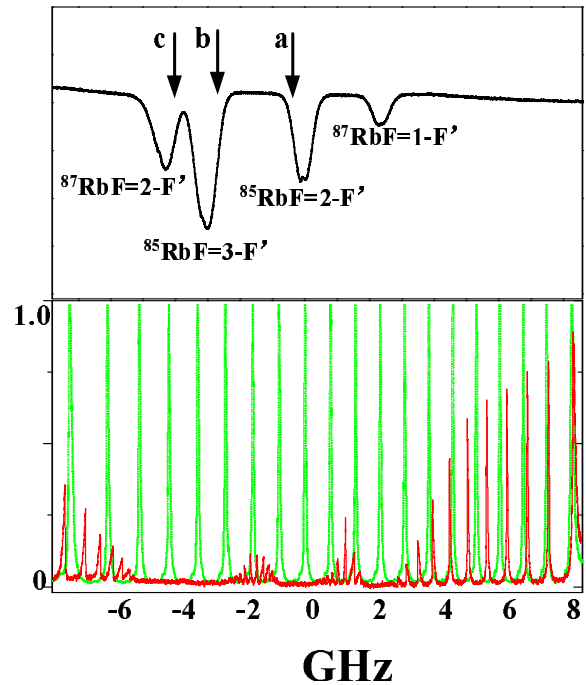}
} \vspace{0.1in}
\caption{(a) The saturated absorption spectrum is shown for the
frequency reference of the pump light. (b) The weak-field cavity
transmission spectrum with atoms inside the cavity (red line). The
empty cavity transmission spectrum without atoms (green line).
\label{Fig2} }
\end{figure}

Figure 3 presents the F-P cavity transmission spectra when the pump
beam frequency is set at different positions as indicated in Fig.
2(a) and the pump, Stokes and anti-Stokes fields build up on
resonance in the cavity simultaneously. First, as the pump frequency
$\omega_{p}$ is tuned to the frequency as indicated by the arrow
$a$, the F-P cavity transmission spectrum is given by Fig. 3(a). The
pump beam power is set at 100 $mW$. The two nearby side peaks
(separated from the large middle peak of the pump light by $\pm6.8$
$GHz$) are from the generated Stokes and anti-Stokes fields of the
$^{87}Rb$ atoms, which oscillate above thresholds. The central peak
and the free-spectral range (FSR) are for the pump beam. The
different peak heights (corresponding to the output power) for the
Stokes and anti-Stokes fields come from several mechanisms including
the cavity detuning, the dependence of gain on the detuning of the
pump field, and the variation of intracavity absorption by other
atomic transitions (as can be easily seen from Fig. 2). The two
outer small peaks on both sides are also the Stokes and anti-Stokes
fields due to the periodicity of the scanned F-P cavity (i.e. FSR).
As the pump frequency is tuned to the position marked by arrow $b$,
the F-P cavity transmission spectrum given by Fig. 3(b) presents the
generated Stokes field to be on resonance with the cavity and above
threshold, where the three-level atomic system is $^{85}Rb$ atoms
with the ground hyperfine state separation of about 3.035 $GHz$. The
anti-Stokes field is absorbed inside the cavity below threshold.
When the pump frequency is further tuned to the red side (arrow c),
both F-P cavity transmission peaks for the Stokes and anti-Stokes
fields appear simultaneously, showing triple-resonant oscillations
for the atom-cavity system. Under both different polarization
configurations of the cavity input field, we can detect the
generated Stokes and anti-Stokes fields with different polarizations
above thresholds.

%
\begin{figure}
\centerline{
\includegraphics[width=3in]{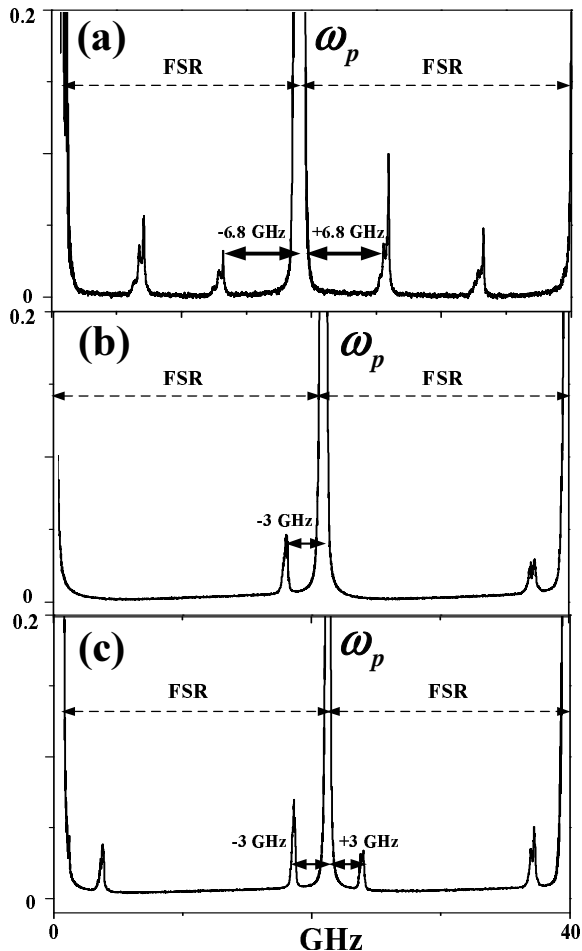}
} \vspace{0.1in}
\caption{The scanned F-P cavity transmission spectra. (a) , (b) and
(c) show the generated Stokes and anti-Stokes fields when the pump
frequency is set at different positions a, b, and c respectively as
indicated in Fig. 2(a). The temperature of the vapor cell is set at
$105^{0}C$. \label{Fig3} }
\end{figure}

The measurement of the oscillation threshold for the anti-Stokes
field (shown in Fig. 3(a)) is presented in Fig.4. The cavity
oscillation starts gradually at low input pump power, and saturates
at high pump power as expected in a laser-like system. The threshold
behaviors for the Stokes and anti-Stokes fields in other cases (for
examples Figs. 3(b) and (c)) are similar. The total output powers of
the Stokes and anti-Stokes fields can reach more than 1 $mW$ when
the pump power is about 100 $mW$, which indicates that such OPO
system with multi-level atoms can be very efficient. The current
double-$\Lambda$ atomic configuration is an ideal system to generate
correlated Stokes and anti-Stokes photon pairs \cite{Lett2}. With
such atomic medium inside an optical cavity and driven above
oscillation thresholds, bright correlated twin beams can be
obtained, similar to the triple-resonant nondegenerate OPOs above
thresholds with $\chi^{(2)}$ nonlinear crystal
\cite{Fabre,Villar,Villar1,Villar2,Coelho,Dauria,Su,Laurat}. The
bright Stokes and anti-Stokes cavity output beams above threshold
will possess the different quantum characteristics comparing with
that produced from single-pass four-wave mixing process with
coherent signal injection \cite{Lett2}. Many interesting phenomena
can be realized using this demonstrated triple-resonant OPO system
in double-$\Lambda$ atomic system, especially with easily managed
large intracavity dispersions.

%
\begin{figure}
\centerline{
\includegraphics[width=3in]{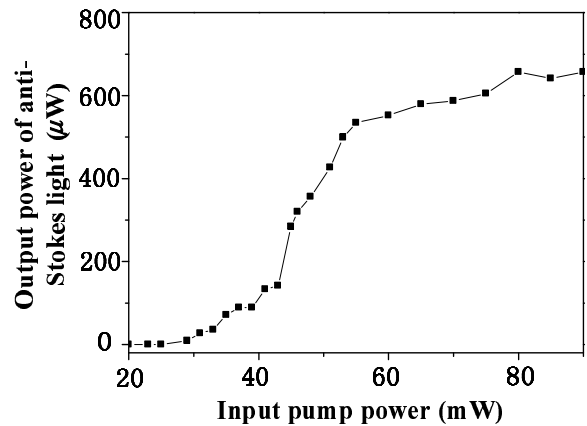}
} \vspace{0.1in}
\caption{The measured threshold behavior of the emitted anti-Stokes
light in Fig. 3(b) as a function the pump beam power. The
temperature of the Rb cell is $T=105^{0}C$. \label{Fig4} }
\end{figure}

In conclusion, we have experimentally demonstrated simultaneous
resonances of the generated Stokes and anti-Stokes fields, together
with the single pump field, in an optical cavity. The modified large
dispersions due to the intracavity dense atomic medium are very
important in allowing such triple-resonant conditions to be
simultaneously satisfied. Such experimental system can be useful to
study atom-cavity interaction, especially for quantum entangled
beams in quantum information processing \cite{app}.

$^{\dagger} $Corresponding author's email address:
jzhang74@yahoo.com, jzhang74@sxu.edu.cn

\smallskip \acknowledgments

We thanks K. Peng, C. Xie and T. Zhang for the helpful discussions.
This research was supported in part by NSFC for Distinguished Young
Scholars (Grant No. 10725416), National Basic Research Program of
China (Grant No. 2006CB921101), NSFC Project for Excellent Research
Team (Grant No. 60821004), and NSFC (Grant No. 60678029). M. X.
acknowledges the funding support from NSF (US).


\end{document}